\documentclass[aps,prl,twocolumn,showpacs,groupedaddress,floatfix,nofootinbib]{revtex4-1}
\usepackage{graphicx}
\usepackage{bm}				
\usepackage{amsfonts}
\usepackage{amssymb}
\usepackage{amsmath}
\usepackage{ulem} 
\usepackage{graphicx}		
\usepackage{xcolor, soul}	
\usepackage{verbatim}
\usepackage[T1]{fontenc}
\usepackage{pifont}
\usepackage{multirow}
\usepackage{notes2bib}
\usepackage{mathrsfs}
\usepackage{bibunits}
\usepackage{appendix}

\begin{document}
\title{Fermi Surface Spin Texture and Topological Superconductivity in Spin-Orbit Free Non-Collinear Antiferromagnets}	
	
\author{Seung Hun Lee$^{1,2,3}$}
\author{Bohm-Jung Yang$^{1,2,3}$}
\email{bjyang@snu.ac.kr}
\affiliation{$^1$Center for Correlated Electron Systems, Institute for Basic Science (IBS), Seoul 08826, Korea\\$^2$Department of Physics and Astronomy, Seoul National University, Seoul 08826, Korea\\$^{3}$Center for Theoretical Physics (CTP), Seoul National University, Seoul 08826, Korea}
	
\date{\today}
	
\begin{abstract}
	We explore the relationship among the magnetic ordering in real space, the resulting spin texture on the Fermi surface, and the related superconducting gap structure in non-collinear antiferromagnetic metals without spin-orbit coupling. Via a perturbative approach, we show that a non-collinear magnetic ordering in a metal can generate a momentum-dependent spin texture on its Fermi surface, even in the absence of spin-orbit coupling, if the metal has more than three sublattices in its magnetic unit cell. Thus, our theory naturally extends the idea of altermagnetism to non-collinear spin structures. 
	When superconductivity is developed in a magnetic metal, as the gap-opening condition is strongly constrained by the spin texture, the nodal structure of the superconducting state is also enforced by the magnetism-induced spin texture.
	Taking the non-collinear antiferromagnet on the kagome lattice as a representative example, we demonstrate how the Fermi surface spin texture induced by noncollinear antiferromagnetism naturally leads to odd-parity spin-triplet superconductivity with nontrivial topological properties.
\end{abstract}

\maketitle

\indent
\textit{Introduction.}---The pairing symmetry of superconductivity is governed by the symmetry which constrains the relative spin directions of electron pairs at opposite momenta on the Fermi surface (FS)~\cite{sigrist1991phenomenological}. Time-reversal $T$ and inversion $P$ symmetries are the representative examples, both of which guarantee the presence of energetically degenerate electrons at opposite momenta while they constrain the spin directions of the electron pairs in an opposite way. When both $T$ and $P$ exist simultaneously, spin-singlet superconductivity which can open a full gap on the spin-degenerate FS is mostly favored. Thus, to achieve unconventional pairing symmetries, such as spin-triplet superconductivity (STS), breaking either $T$ or $P$ is normally considered.

For instance, various unconventional superconducting states are proposed in $T$-symmetric metals with broken $P$ symmetry~\cite{potter2011engineering,smidman2017superconductivity,li2019exploring,desjardins2019synthetic,yoshizawa2021atomic,zhang2022topological}. In such noncentrosymmetric metals, as parity-mixing occurs due to $P$-breaking, superconducting states can contain spin-triplet components. Moreover, when spin-orbit coupling (SOC) is present, the $T$-symmetric spin-split FS can host a spin-momentum locked winding spin texture, which leads to STS with intriguing nodal structure~\cite{smidman2017superconductivity}.
Because of the Kramers' degeneracy at $T$-invariant momenta, such winding spin texture carries a topological charge and appears robustly. 
However, unless the superconducting state is dominated by spin-triplet components, observing nodal STS is not easy to achieve. 

On the other hand, in centrosymmetric magnetic metals preserving $P$ but breaking $T$, equal-spin triplet pairing can naturally arise because $P$ enforces electron pairs at opposite momenta to have the same spin direction. 
Thus, if the FS is spin-split, $P$-symmetric magnetic metals are promising candidates to achieve STS.
Interestingly, as shown in Ref.~\cite{lee2021odd}, such a spin-split FS with fixed spin polarization can appear not only in ferromagnets but also in collinear antiferromagnets when $P$ is broken locally but preserved globally~\cite{vsmejkal2020crystal,lee2021odd,vsmejkal2022beyond,vsmejkal2022emerging}. Thus, centrosymmetric collinear magnetic metals can also host odd-parity STS. However, in such systems, the FS generally does not have robust spin texture even when SOC exists. This is because, as the Kramers' degeneracy is lifted due to broken $T$, winding spin texture with topological stability does not appear in general.
Therefore, to achieve nodal STS in centrosymmetric magnets, a distinct mechanism to protect winding spin texture is necessary.

In this letter, we study how real space magnetic ordering of a centrosymmetric non-collinear antiferromagnetic (AFM) metal induces winding spin texture on the FS without SOC, which in turn constrains the gap structure of its superconducting state. Through a perturbation approach, we show that if the number of sublattices in a magnetic unit cell is larger than two, the FS can have momentum-dependent, winding spin texture. Thus, our theory extends the idea of altermagnetism proposed recently \cite{vsmejkal2022beyond,vsmejkal2022emerging} to the cases of non-collinear antiferromagnets without SOC.
We demonstrate our theory in a kagome non-collinear antiferromagnet with three sublattices. 
As a result of magnetism-induced winding spin texture on the FS, we find various types of topological superconductors (TSCs) with the odd-parity spin-triplet pairing, including nodal TSCs, and first-order and second-order TSCs.

\indent
\textit{Magnetic ordering and FS spin texture.}---To model a generic AFM metal with $n$ sublattices in its magnetic unit cell, we construct a tight-binding Hamiltonian $\hat{H}=\sum_{\bm{k}}\hat{\bm{c}}_{\bm{k}}^{\dagger}\mathcal{H}(\bm{k})\hat{\bm{c}}_{\bm{k}}$ where we take the basis $\hat{\bm{c}}_{\bm{k}}^{\dagger}=(\hat{c}_{\bm{k}1\uparrow}^{\dagger},\hat{c}_{\bm{k}1\downarrow}^{\dagger},\hat{c}_{\bm{k}2\uparrow}^{\dagger},\hat{c}_{\bm{k}2\downarrow}^{\dagger},\cdots,\hat{c}_{\bm{k}n\uparrow}^{\dagger},\hat{c}_{\bm{k}n\downarrow}^{\dagger})$.
When $n=2$, we have 
\begin{equation}\label{n2}
\mathcal{H}(\bm{k})=
\begin{pmatrix}
h_{11}(\bm{k})	& h_{12}(\bm{k})\\
h_{21}(\bm{k})	& h_{22}(\bm{k})
\end{pmatrix},
\end{equation}
where $h_{ij}(\bm{k})$ ($i,j=1,2$) are $2\times2$ block Hamiltonians. The diagonal blocks $h_{ii}(\bm{k})=-\bm{m}_i\cdot\bm{\sigma}$ describe mean-field approximated local spin orders represented by a constant vector $\bm{m}_i$, while the off-diagonal blocks $h_{12}(\bm{k})=h_{\text{NN}}(\bm{k})\sigma_0=h_{21}^\dagger(\bm{k})$ describe the kinetic part of the Hamiltonian coming from the nearest-neighbor (NN) hopping between different sublattices. 

The Green's function of $\mathcal{H}(\bm{k})$ is given by $G(\bm{k},\varepsilon)=[\varepsilon I_{4\times4}-\mathcal{H}(\bm{k})]^{-1}$, where $I_{l\times l}$ is the identity matrix of size $l$~\cite{kawano2019discovering}. From $g_{ii}(\bm{k},\varepsilon)$, the $2\times2$ diagonal blocks of $G(\bm{k},\varepsilon)$, we extract the effective Hamiltonian $h_{\text{eff}}^{i}(\bm{k})$ projected onto the $i$-th sublattice by using the equation $g_{ii}(\bm{k},\varepsilon)=[\varepsilon I_{2\times2}-h_{\text{eff}}^{i}(\bm{k})]^{-1}$. For instance, we obtain
\begin{align}\label{n2heff1coll}
h_{\text{eff}}^{1}(\bm{k})&=h_{11}(\bm{k})+h_{12}(\bm{k})[\varepsilon I_{2\times2}-h_{22}(\bm{k})]^{-1}h_{21}(\bm{k})\nonumber\\
&=-\bm{m}_1\cdot\bm{\sigma}+|h_{\text{NN}}(\bm{k})|^2[\varepsilon I_{2\times2}+\bm{m}_2\cdot\bm{\sigma}]^{-1}\nonumber\\
&=-\bm{m}_1\cdot\bm{\sigma}+\frac{|h_{\text{NN}}(\bm{k})|^2}{m_2^2-\varepsilon^2}[\varepsilon I_{2\times2}-\bm{m}_2\cdot\bm{\sigma}].
\end{align}
for $i=1$.

\begin{figure}[t!]
	\centering
	\includegraphics[width=1\linewidth]{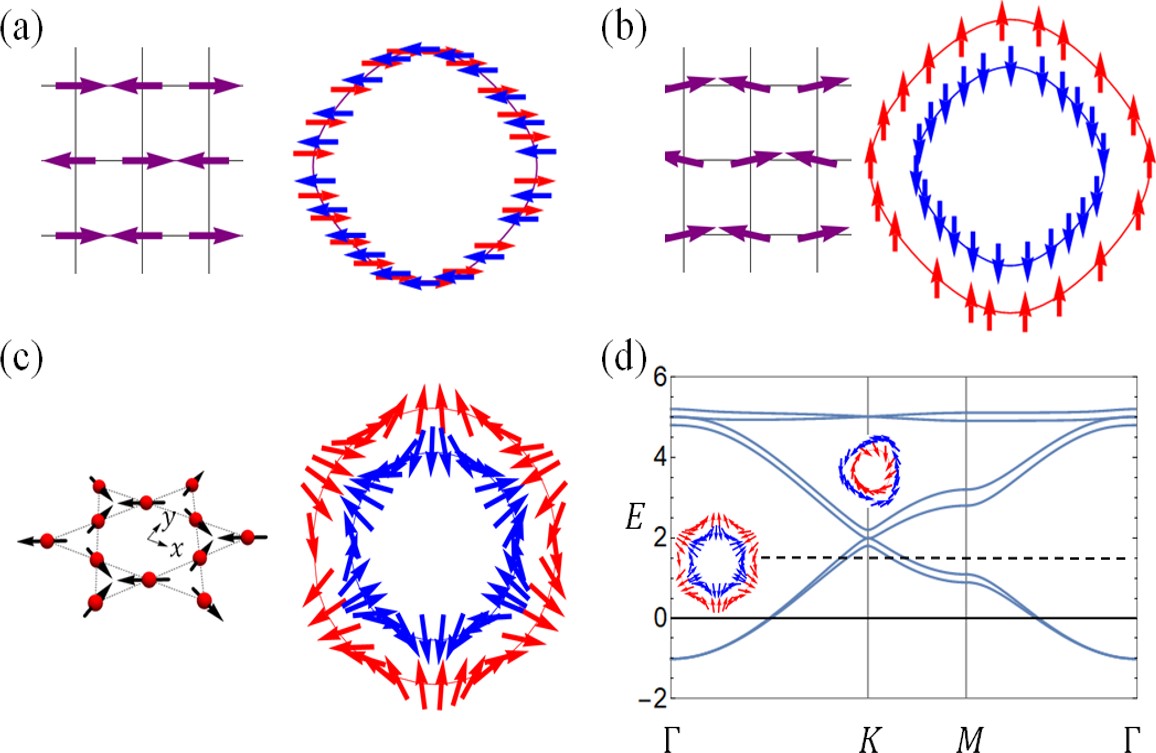}
	\caption{(a) The FSST of a collinear antiferromagnet in the square lattice $(\bm{m}_1=-\bm{m}_2)$. The FS is spin degenerate while the spin direction is parallel to the AFM ordering direction.
		(b) The FSST of a canted antiferromagnet ($\bm{m}_1+\bm{m}_2\neq0$). The FS is spin-split while the spin direction on each FS is uniform.
		(c) The kagome AIAO AFM structure (left) and the corresponding FSST (right) computed when $m=0.2$ and $\mu=-3.0$.
		(d) The energy eigenvalues of $\mathcal{H}_{\text{KAFM}}(\bm{k})$ calculated along high-symmety momentum directions. 
		The FSST for $\mu=-3.0$ ($-1.5$) at which the FS encloses the $\Gamma$ ($K$ and $K'$) is also plotted. The solid (dotted) black horizontal line indicates the Fermi level at $\mu=-3.0$ ($-1.5$).
	}\label{fig1}
\end{figure}

If we separate the effective Hamiltonian into spin-independent and spin-dependent parts as $h_{\text{eff}}^{1}(\bm{k})=R_0^1(\bm{k})\sigma_0+\bm{R}^1(\bm{k})\cdot\bm{\sigma}$, Eq.~(\ref{n2heff1coll}) gives $\bm{R}^1(\bm{k},\varepsilon)=-[\bm{m}_1+\bm{m}_2|h_{\text{NN}}(\bm{k})|^2/(m_2^2-\varepsilon^2)]$.
Since the FS is an equi-energy contour, $|h_{\text{NN}}(\bm{k})|^2$ and $\varepsilon$ are constant on the FS. Hence $\hat{R}^1(\bm{k},\varepsilon)\equiv\bm{R}^1(\bm{k},\varepsilon)/|\bm{R}^1(\bm{k},\varepsilon)|$ is uniform on the FS for a given Fermi level as shown in Fig.~\ref{fig1} (a) and (b). One can see that when $\bm{m}_2\nparallel\bm{m}_1$, the correction term from $h_{22}(\bm{k})$ in Eq.~(\ref{n2heff1coll}), proportional to $\bm{m}_2\cdot\bm{\sigma}$, tilts $\hat{R}^1(\bm{k},\varepsilon)$ away from (towards) $\bm{m}_1$ ($\bm{m}_2$).

In the presence of additional terms such as the next-nearest-neighbor (NNN) hopping which enters $h_{11}(\bm{k})$ and $h_{22}(\bm{k})$ in the form of $h_{\text{NNN}}(\bm{k})\sigma_0$, the direction of $\bm{R}^1(\bm{k},\varepsilon)$ may depend on $\bm{k}$. Nevertheless, $\bm{m}_1$ and $\bm{m}_2$ cannot generate smooth `winding' FS spin texture (FSST) when $n=2$. This is because, to achieve winding FSST, there should be a momentum $\bm{k}'$ that satisfies $\hat{R}^1(\bm{k}',\varepsilon)=-\hat{R}^1(\bm{k},\varepsilon)$ for an arbitrary $\bm{k}$. However, this condition can never be fulfilled unless $\bm{m}_1$ and $\bm{m}_2$ are either parallel or antiparallel. If they are parallel or antiparallel, Eq.~(\ref{n2heff1coll}) gives $\hat{R}^1(\bm{k}',\varepsilon)\parallel\bm{m}_1\parallel\bm{m}_2$. The recently proposed altermagnets belong to this case~\cite{vsmejkal2022beyond,vsmejkal2022emerging}.

On the other hand, when $n\geq3$, the FS spin direction can be momentum-dependent in general. For $n=3$, we obtain $\bm{R}^1(\bm{k},\varepsilon)=-\bm{m}_1-\mathfrak{f}(\bm{k})\bm{m}_2-\mathfrak{g}(\bm{k})\bm{m}_3-\mathfrak{h}(\bm{k})(\bm{m}_2\times\bm{m}_3)$, where $\mathfrak{f}(\bm{k})$, $\mathfrak{g}(\bm{k})$ and $\mathfrak{h}(\bm{k})$ are $\bm{k}$-dependent functions whose explicit forms are provided in Supplementary Material (SM)~\cite{supplement}. 
Since $\bm{R}^i(\bm{k})$ for a given $\varepsilon$ is a real vector field in the momentum space, it can have a winding number $w$ on the FS if i) $\bm{R}^i(\bm{k})$ is confined in a two-dimensional plane and ii) the FS is a closed curve enclosing singular points $\bm{k}_{c=1,2,\cdots,n_c}$ where $\bm{R}^i(\bm{k}_{c},\varepsilon)=0$.
Then, $w=\sum_{c=1}^{n_c}v_c$ where $v_c$ is the vorticity of the $c$-th singular point. This condition can be satisfied with an appropriate choice of $h_{ij}(\bm{k})$ terms. Notably, the presence of $\mathfrak{h}(\bm{k})$ term, which originates from the higher-order correction term, proportional to $(\bm{m}_2\cdot\bm{\sigma})(\bm{m}_3\cdot\bm{\sigma})$, indicates that the dimension of $\bm{R}^i(\bm{k})$ can be higher than that $\bm{m}_{i=1,2,3}$ can span in general~\cite{xiao2023spin}. However, if it is possible to choose the real gauge for the kinetic part of the Hamiltonian owing to symmetries such as $C_{2z}T$ in two-dimension (2D) where $C_{m\alpha}$ is $m$-fold rotation about the $\alpha$-axis or $PT$ in two- or three-dimension (3D), $\mathfrak{h}(\bm{k})$ vanishes for a generic $\bm{k}$~\cite{supplement,ahn2018band,ahn2019failure}. Because of this, especially in 2D, the winding number of FSST is well-defined only when i) $C_{2z}T$ or $PT$ is present (so $\mathfrak{h}(\bm{k})=0$) and ii) $\bm{m}_{i=1,2,3}$ are linearly dependent, or equivalently, $\bm{m}_{i=1,2,3}$ are coplanar but not parallel to each other. The kagome non-collinear antiferromagnet is an example that satisfies the above conditions.

\indent
\textit{Kagome non-collinear antiferromagnet}.---
As an example demonstrating the relation among the FSST, real space magnetic ordering, and pairing symmetries~\cite{zhang2022topological,almeida2017induced,powell2003gap,lee2021odd}, we consider the kagome lattice non-collinear antiferromagnet whose structure is intrinsically $P$-symmetric globally but $P$-asymmetric locally.

The ground state of the classical Heisenberg AFM with Dzyaloshinskii-Moriya interaction on the kagome lattice is known to have coplanar $120^\circ$ ordering lying in the lattice plane, in which the angles between NN spins are all $120^\circ$~\cite{sachdev1992kagome}. Among the infinitely degenerate coplanar $120^\circ$ ordered states, we choose the so-called all-in-all-out (AIAO) order (Fig.~\ref{fig1} (c)) that has three in-plane two-fold rotation symmetries.

We construct a tight-binding Hamiltonian $\mathcal{H}_{\text{KAFM}}(\bm{k})$ describing the kagome AIAO antiferromagnet as in Eq.~(\ref{n2}) but with $n=3$. Explicitly, we have $h_{12}=t(1+e^{i\bm{k}\cdot\bm{e}_1})$, $h_{13}=t(1+e^{-i\bm{k}\cdot\bm{e}_3})$, $h_{23}=t(1+e^{i\bm{k}\cdot\bm{e}_2})$, and $h_{ii}(\bm{k})=-\mu-\bm{m}_i\cdot\bm{\sigma}$ where $\bm{e}_1=(1,0)$, $\bm{e}_2=(-1/2,\sqrt{3}/2)$, $\bm{e}_3=(-1/2,-\sqrt{3}/2)$, $\bm{m}_1=m(-\sqrt{3}/2,-1/2)$, $\bm{m}_2=m(\sqrt{3}/2,-1/2)$, and $\bm{m}_3=m(0,1)$. Here, $t=1$, $m$, and $\mu$ indicate the NN hopping amplitude, local magnetic moment, and chemical potential, respectively. In Fig.~\ref{fig1} (d), we show the corresponding band structure for $m=0.2$ and $\mu=-3.0$.
The $\bm{R}^i$ field of the kagome AIAO AFM has singular points with total vorticity $-2$ (+1) near the $\Gamma$ point ($K$ and $K'$ points). Therefore, the two Fermi pockets that enclose $\Gamma$ carry $w=-2$ FSST as in Fig.~\ref{fig1} (c), while the pockets that enclose $K$ or $K'$ carry $w=1$. The relevant FSSTs are also shown in Fig.~\ref{fig1} (d).
This example clearly demonstrates that FSST can be generated by real space magnetic ordering in the absence of SOC~\cite{hayami2020bottom,hayami2020spontaneous,mcclarty2023landau}.

In addition, the vanishing $\bm{R}^i$ field at the singular points implies possible two-fold degeneracies of spin bands. In Fig.~\ref{fig1} (d), among the nodes at $C_{3z}$-invariant momenta protected by $C_{3z}$ and $C_{2z}T$ symmetries, the gap closings between first and second lowest bands at $\Gamma$ and the fifth and sixth lowest bands at $K$ ($K'$) are such nodes. We note that the band touching between the second and third lowest bands at $K$ ($K'$) shows quadratic dispersion, even though the FSST winds only once. This happens because the physical spin (winding once) is different from the pseudospin (winding twice) defined in the projected two-band space related to the band crossing.

\indent
\textit{FSST and superconducting gap structure.}---
The spin structure of spin-triplet Cooper pairs can be described by the so-called $d$-vector. Thus, the effect of the FSST on Cooper pairs can be understood from the relation between the $d$-vector and the FS spin direction. 
With Pauli matrices $\sigma_{x,y,z}$ that act on spin degrees of freedom, the mean-field pairing interaction can be written in a simple form $\hat{H}_{\text{int}}=\sum_{\bm{k},s,s'}\Delta_{\bm{k}}\hat{c}_{\bm{k},s}^{\dagger}\hat{c}_{-\bm{k},s'}^{\dagger}+\text{h.c.}$, where $\Delta_{\bm{k}}=\psi_0(\bm{k})(i\sigma_y)+\bm{d}(\bm{k})\cdot\bm{\sigma}(i\sigma_y)$ and $s$ ($s'$) denotes the spin. For a given spin basis, the $d_z$ component indicates opposite-spin triplet pairing, while the $d_x$ and $d_y$ components describe equal-spin triplet pairing. 
As shown in Fig.~\ref{fig2} (a), if $P$ is present while $T$ is absent, the equal-spin Cooper pairing is forced in the weak-coupling limit, because electron pairs at opposite momenta have the same spins. Therefore, for the spin basis aligned to the FSST direction, $\bm{d}(\bm{k})$ should have a component perpendicular to the FSST to open a gap on the FS. On the other hand, if $T$ is present while $P$ is absent, the opposite-spin Cooper pairing is forced, so the pairing interaction should have a $\bm{d}(\bm{k})$ component that is parallel to the FSST to open a gap.

\begin{figure}[t!]
	\centering
	\includegraphics[width=1\linewidth]{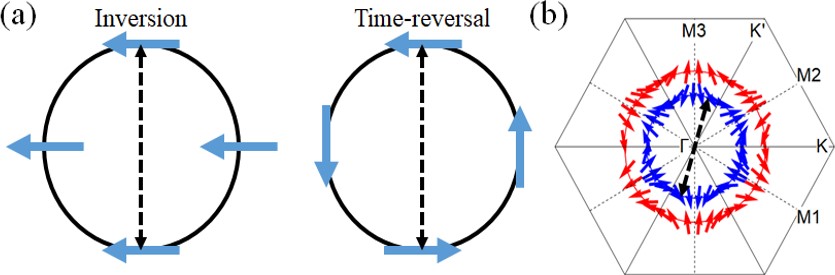}
	\caption{(a) Schematic diagram describing the FS spin direction with $P$ (left) and $T$ (right) symmetries.
		(b) The FSST of $\mathcal{H}_{\text{eff}}$ that corresponds to Fig.~\ref{fig1} (c). The in-plane two-fold rotation axes and the mirror invariant lines are represented by dashed and solid lines, respectively. The black dashed arrow connects the electron pair at opposite momenta for each case.
	}\label{fig2}
\end{figure}

To describe the superconducting state of the kagome antiferromagnet, we introduce pairing interaction $\Delta(\bm{k})$ and rewrite the Hamiltonian in the Nambu basis $\hat{\Psi}_{\bm{k}}^{\dagger}=(\hat{\bm{c}}_{\bm{k}}^{\dagger},\hat{\bm{c}}_{-\bm{k}})$. Then, the Bogoliubov-de Gennes (BdG) Hamiltonian becomes $\sum_{\bm{k}}\hat{\Psi}_{\bm{k}}^{\dagger}\mathcal{H}_{\textrm{BdG}}(\bm{k})\hat{\Psi}_{\bm{k}}$ with
\begin{equation}\label{eq:BdG}
\mathcal{H}_{\text{BdG}}(\bm{k})=
\begin{pmatrix}
\mathcal{H}_{\text{KAFM}}(\bm{k})-\mu	& \Delta(\bm{k})\\
\Delta^{\dagger}(\bm{k})	& -\mathcal{H}_{\text{KAFM}}^T(-\bm{k})+\mu
\end{pmatrix}.
\end{equation}

\indent
Since $\mathcal{H}_{\text{KAFM}}(\bm{k})$ is symmetric under $D_{3d}$ point group symmetries, the pairing channels can be classified by corresponding irreducible representations (IRs)~\cite{sigrist1991phenomenological}. As the AIAO AFM state is $P$-symmetric but $T$-broken, equal-spin triplet pairing is favored, and thus only odd-parity spin-triplet IRs ($A_{1u}$, $A_{2u}$, and $E_u$) can induce superconducting instability within the weak-coupling approximation. The transformation properties of the IRs under the $D_{3d}$ point group are summarized in SM~\cite{supplement}.

To illustrate the relation between the gap structure and FSST, we consider the FS with $w=2$ near $\Gamma$ shown in Fig.~\ref{fig1} (d) as an example.
Projecting the Hamiltonian $\mathcal{H}$ onto the lowest energy band of the nonmagnetic kagome lattice Hamiltonian $\mathcal{H}_{\text{KAFM}}(\bm{k})|_{\bm{m}_i=0}$ and expanding it up to the quadratic order of $k_x$ and $k_y$, we obtain the effective Hamiltonian near $\Gamma$ given by
\begin{equation}\label{eff}
\mathcal{H}_{\text{eff}}(\bm{k})=(-4+k_x^2+k_y^2)+\frac{m}{6}k_xk_y\tau_x+\frac{m}{12}(k_x^2-k_y^2)\tau_y,
\end{equation}
where the Pauli matrices $\tau_{i=1,2,3}$ denote the effective spin that represents the lowest two bands in Fig.~\ref{fig1} (d).  For $\mathcal{H}_{\text{eff}}(\bm{k})$, the matrix representations of the $D_{3d}$ symmetry generators are given by $C_{3z}=(\tau_0-i\sqrt{3}\tau_z)/2$, $C_{2y}=i\tau_y$, and $P=\tau_0$.


Let us first consider $\Delta_{\bm{k}}^{A_{1u}}$, the pairing interaction belonging to the $A_{1u}$ IR. As it satisfies $C_{2y}(\bm{k})\Delta_{\bm{k}}^{A_{1u}}(\bm{k})C_{2y}^T(\bm{k})=\Delta_{C_{2y}\bm{k}}^{A_{1u}}$, $d_x$ and $d_z$ components, which couple to $\tau_x(i\tau_y)$ and $\tau_z(i\tau_y)$ in $\Delta_{\bm{k}}^{A_{1u}}$, respectively, are forbidden on the $k_y$-axis. As displayed in Fig.~\ref{fig2} (b), the effective spin texture on the $k_y$-axis is pointing to the $y$-direction, parallel to the $d$-vector of $\Delta_{\bm{k}}^{A_{1u}}$, thus $\Delta_{\bm{k}}^{A_{1u}}$ cannot open a gap on the $k_y$ axis. Similar arguments can also be applied to two other in-plane two-fold rotation axes. Thus, $\Delta_{A_{1u}}$ cannot open a gap at the intersection of the FS and the $\Gamma-M_{i=1,2,3}$ lines.

On the other hand, $\Delta_{\bm{k}}^{A_{2u}}$ satisfies $M_y(\bm{k})\Delta_{\bm{k}}^{A_{2u}}(\bm{k})M_y^T(\bm{k})=\Delta_{M_y\bm{k}}^{A_{2u}}$ where $M_y=PC_{2y}$ is a mirror reflection with respect to the $zx$-plane. Thus, the $d$-vector of $\Delta_{\bm{k}}^{A_{2u}}$ is parallel to the spin texture on the $k_x$-axis, both aligned to the $y$-direction. As a result, $\Delta_{\bm{k}}^{A_{2u}}$ has nodes along the mirror invariant lines ($\Gamma-K$, $\Gamma-K'$, and $K-K'$).

Finally, in the case of the $E_{u}$ IR, there is no such constraints for the direction of the $d$-vector (see SM~\cite{supplement}).

\begin{figure}[t!]
	\centering
	\includegraphics[width=1\linewidth]{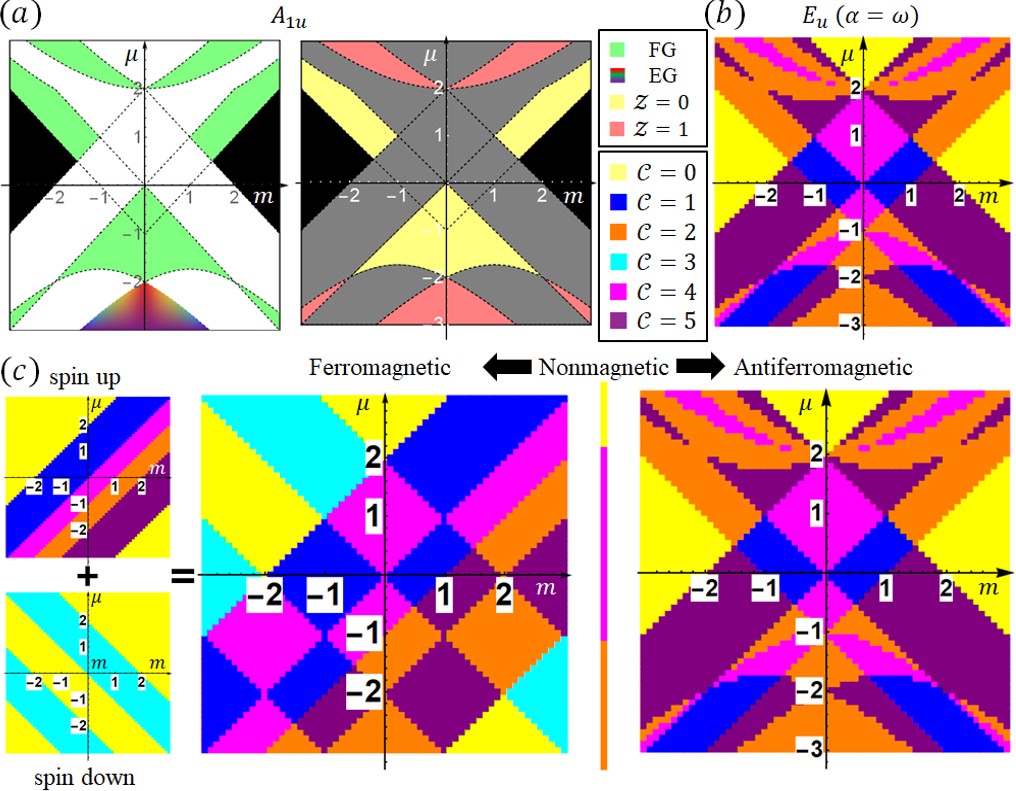}
	\caption{
		(a) The topological phase diagrams for the $A_{1u}$ IR obtained in the weak pairing limit. In the black-colored region, the normal state is insulating. In the white-, green-, and rainbow-colored regions, the $A_{1u}$ pairing induces nodal, fully gapped (FG), and gapless but easily gappable (EG) superconductivity, respectively. The yellow- and red-colored regions in the right figure represent the trivial superconductor ($\mathcal{Z}=0$) and the second-order TSC ($\mathcal{Z}=1$), respectively.
		(b) The topological phase diagrams for the $E_{u}$ IR with $C_{3z}$ eigenvalue $\alpha=\omega$. The yellow, blue, orange, cyan, magenta, and purple regions represent the phases with $\mathcal{C}=0$, $\mathcal{C}=1$, $\mathcal{C}=2$, $\mathcal{C}=3$, $\mathcal{C}=4$, and $\mathcal{C}=5$ modulo 6, respectively. For $\alpha=\omega^2$, the sign of $\mathcal{C}$ becomes reversed. 
		(c) The evolution of the phase diagram from the nonmagnetic case to the ferromagnetic case (left) and the non-collinear AFM case in (b) (right). To consider a situation where both $P$ and $C_{3z}$ are conserved, we assume out-of-plane ordered ferromagnetism in the left. Then, the ferromagnetic phase diagram can be simply obtained by superposing spin-up and -down phase diagrams, while the AFM phase diagram shows a more complex structure.
	}\label{fig3}
\end{figure}

\indent
\textit{Topological superconductivity.}---
Consistent with the gap structure analysis by comparing the $d$-vector direction and FSST, one can find that the superconducting state is always gapped for the $E_u$ pairing. Meanwhile, for the $A_{1u}$ or $A_{2u}$ pairing, the bulk is gapless if the FS intersects high-symmetry lines where the $d$-vector is parallel to the FSST, unless the FS avoids such high-symmetry lines. 
To reveal the nature of the superconducting states in the kagome antiferromagnet,
we investigate the nodal and topological structures of the BdG Hamilitonian in Eq.~(\ref{eq:BdG}) in weak pairing limit by varying $\mu$ and $m$, as summarized in Fig.~\ref{fig3}.
We note that the $A_{2u}$ pairing makes the system gapless with point nodes on the mirror-invariant lines in every region of the $m-\mu$ parameter space where the normal state is metallic. 
In contrast, while the $A_{1u}$ pairing mostly induces gapless superconducting states with nodes along the $C_2$-invariant lines, it can also induce a fully gapped superconductor in the green-colored regions in Fig.~\ref{fig3} (a) where the FS does not cross the $C_2$-invariant lines.
In both the $A_{1u}$ and $A_{2u}$ pairings, as the nodes are located along high-symmetry lines, pair-annihilation between neighboring nodes can sometimes happen when the pairing interaction is strong enough.
The resulting gapped superconducting state can become a second-order TSC as shown in the red region in Fig.~\ref{fig3} (a).

In the case of gapped superconducting states, its topological properties can be described by using the symmetry indicators~\cite{skurativska2020atomic,ono2020refined,ono2021z,elcoro2021magnetic}.
When the pairing interaction is weak enough, the symmetry representations of the occupied states of $\mathcal{H}_{\text{BdG}}(\bm{k})$ at high-symmetry momenta remain unchanged after the superconducting phase transition. Thus, regardless of the detailed form of the pairing interaction, one can draw the phase diagrams using the symmetry indicator information extracted from the normal state and pairing symmetry as discussed in detail in SM~\cite{supplement}.
The distribution of the invariant $\mathcal{Z}$ related to the second-order topology for the $A_{1u}$ IR and that of the Chern number for the $E_{u}$ IR, obtained by using the symmetry indicators, are shown in Fig.~\ref{fig3} (a) and (b), respectively. The corresponding Wilson loop and edge spectra are described in SM~\cite{supplement}.

We note that the spin texture induced by noncollinear AFM ordering also affects the topological nature of the gapped superconducting states. To illustrate this, we compare the superconducting phase diagram of the kagome ferromagnet and that of the noncollinear antiferromagnet considering the $E_{u}$ IR, as shown in Fig.~\ref{fig3} (c).
In the case of the ferromagnet without winding spin texture, the superconducting phase diagram can be understood by using the relevant phase diagram for spinless band structure with $m=0$. Since the ferromagnetism only induces a rigid shift of spin-up and spin-down bands, the superconducting phase diagram of the ferromagnet is merely a simple superposition of two $m=0$ phase diagrams for spin-up and spin-down electrons, respectively.
On the other hand, in the case of noncollinear antiferromagnet, as the winding spin texture accompanies normal state band degeneracies associated with the singular points of the $\bm{R}$ field, which are absent in ferromagnet, the corresponding superconducting phase diagram exhibits distinct topological characteristics.

\textit{Discussion.}---
In this letter, we have studied the effect of real space magnetic texture on the FS and its superconductivity. 
Especially, it is shown that spin-momentum locking on the FS can be generated purely by the real space magnetism even in the absence of SOC. This magnetism-induced FSST, together with the $d$-vector, provides an intuitive understanding of the symmetry-protected nodes on high-symmetry lines in non-collinear AFM superconductors.
We emphasize that the scope of our theory is not restricted to the kagome lattice in 2D. It can be applied to a general two- or three-dimensional lattice with an arbitrary number of sublattices. Indeed, consistent with our prediction on the requirements to have winding spin texture, a recent spin-resolved angle-resolved photoemission spectroscopy measurement reported that MnTe$_2$~\cite{zhu2023observation}, a non-coplanar antiferromagnet that has four atoms in the magnetic unit cell with linearly dependent local spin moments, shows winding spin texture.
Finally, considering our prediction of various odd-parity spin-triplet superconducting states with intriguing nodal structures and topological properties, we expect that the kagome AFM is poised to become a crucial arena for Majorana engineering in magnetic superconductors.

	\begin{acknowledgments}
		S.H.L. and B.-J.Y. were supported by the Institute for Basic Science in Korea (Grant No. IBS-R009-D1),
		Samsung  Science and Technology Foundation under Project Number SSTF-BA2002-06,
		the National Research Foundation of Korea (NRF) grant funded by the Korean government (MSIT) (No.2021R1A2C4002773, and No. NRF-2021R1A5A1032996). 
	\end{acknowledgments}
	
	\bibliographystyle{apsrev4-1}
	\bibliography{KLAFM}
	
	\newpage
	
	\onecolumngrid
	
	\appendix
	
	\begin{bibunit}
		\renewcommand{\appendixpagename}{\center\large Supplementary Material: Fermi Surface Spin Texture and Topological Superconductivity in Spin-Orbit Free Non-Collinear Antiferromagnets}
		
		\appendixpage
		
		\setcounter{page}{1}
		\setcounter{equation}{0}
		\setcounter{figure}{0}
		\setcounter{table}{0}
		\setcounter{equation}{0}
		
		\renewcommand{\theequation}{S\arabic{equation}}
		\renewcommand{\thetable}{S\arabic{table}}
		\renewcommand{\thefigure}{S\arabic{figure}}
		
		\author{Seung Hun Lee$^{1,2}$}
		\email{sh2lee@snu.ac.kr}
		\author{Bohm-Jung Yang$^{1,2,3}$}
		\email{bjyang@snu.ac.kr}
		\affiliation{$^1$Center for Correlated Electron Systems, Institute for Basic Science (IBS), Seoul 08826, Korea\\$^2$Department of Physics and Astronomy, Seoul National University, Seoul 08826, Korea\\$^{3}$Center for Theoretical Physics (CTP), Seoul National University, Seoul 08826, Korea}
		
		\date{\today}
		
		\maketitle
		
		\noindent
		\textbf{Outline}\\
		\indent In this supplementary material, we provide details of discussions made in the main text. The sections are organized as follows.\\
		\\
		\indent \textbf{Fermi surface spin texture} \hfill \pageref{section1}\\
		\\
		\indent \textbf{Application to two-dimensional metallic antiferromagnets} \hfill \pageref{section2}\\
		\\
		\indent \textbf{$E_u$ irreducible representation} \hfill \pageref{section3}\\
		\\
		\indent \textbf{Topological superconductivity} \hfill \pageref{section4}\\
		\\
		\\
		\noindent
		\textbf{Fermi surface spin texture}\label{section1}\\
		\indent In this section, we show the derivation of the Fermi surface (FS) spin texture equation Eq. (2). We start from the tight-binding Hamiltonian.
		When the Bloch Hamiltonian of electrons in solids with a single orbital per sites is given by
		\begin{equation}
		H(\bm{k})=R_{0}(\bm{k})\sigma_{0}+\bm{R}(\bm{k})\cdot\bm{\sigma}
		\end{equation}
		where $\bm{R}(\bm{k})=(R_{x}(\bm{k}),R_{y}(\bm{k}),R_{z}(\bm{k}))$, $R_{\mu}(\bm{k})\in\mathbb{R}$, and $\sigma_{\mu}$s represent the spin Pauli matrices, $\bm{R}(\bm{k})$ can be regarded as an momentum dependent Zeeman field~\cite{kawano2019discovering}. A non-trivial spin texture on Fermi surfaces can arise due to this effective Zeeman field.\\
		\indent One can extend this picture to multi-degree-of-freedom (orbital, sublattice, etc.) systems by applying Brillouin-Wigner formalism. The Bloch Hamiltonian of an $n$-degree-of-freedom system is described by a $2n\times2n$ ($n>2$) Hermitian matrix. Let us consider $n=2$ case first. The Hamiltonian and its retarded Green function can generally be written as
		\begin{align}
		&H(\bm{k})=\sum_{\mu,\nu=0,x,y,z}h_{\mu\nu}(\bm{k})\tau_{\mu}\otimes\sigma_{\nu}=
		\begin{pmatrix}
		H_{11}(\bm{k})	& H_{12}(\bm{k}) \\
		H_{21}(\bm{k})	& H_{22}(\bm{k})
		\end{pmatrix} \\
		&G(\bm{k},\varepsilon)=[\varepsilon\tau_{0}\otimes\sigma_{0}-H(\bm{k})]^{-1}=
		\begin{pmatrix}
		G_{11}(\bm{k})	& G_{12}(\bm{k}) \\
		G_{21}(\bm{k})	& G_{22}(\bm{k})
		\end{pmatrix}
		\end{align}
		where $h_{\mu\nu}(\bm{k})$, $\varepsilon\in\mathbb{R}$ and $\tau_{\mu}$s denote the Pauli matrices acting on the extra degree-of-freedom space. The poles and residues of $G(\bm{k})$ give all the eigenvalues and eigenstates of $H(\bm{k})$. $G_{ii}(\bm{k})$ ($i=1,2,\cdots,n$), the Green function in the $i$-degree-of-freedom subspace, also gives all the eigenvalues and eigenstates projected onto the $i$-subspace. For $n=2$, we obtain the effective Hamiltonian for $i=1$ subspace as
		\begin{equation}
		G_{11}(\bm{k},\varepsilon)=[\varepsilon\sigma_{0}-H_{\textrm{eff}}^{i=1}(\bm{k})]^{-1}
		\end{equation}
		where
		\begin{align}
		H_{\textrm{eff}}^{i=1}(\bm{k})&=H_{11}(\bm{k})+H_{12}(\bm{k})[\varepsilon\sigma_{0}-H_{22}(\bm{k})]^{-1}H_{21}(\bm{k})\nonumber\\
		&=R^{i=1}_{0}(\bm{k},\varepsilon)+\bm{R}^{i=1}(\bm{k},\varepsilon)\cdot\bm{\sigma}.
		\end{align}
		\\
		\noindent
		\textbf{Application to two-dimensional metallic antiferromagnets}\label{section2}\\\\
		\indent\textit{Collinear antiferromagnets on two-dimensional square lattices}.---
		First, let us construct a tight binding model for a collinear antiferromagnet on a two-dimensional square lattice. We put up to the nearest neighbor hopping into the model and take into account the exchange interaction via mean field approximation. We note that in this section we use $H_{ii}(\bm{k})=\bm{m}_i\cdot\bm{\sigma}$ to describe the magnetic Hamiltonian instead of the notation $h_{ii}(\bm{k})=-\bm{m}_i\cdot\bm{\sigma}$ that used in the main text, but the final results ($\mathfrak{f}(\bm{k})$, $\mathfrak{g}(\bm{k})$, and $\mathfrak{h}(\bm{k})$) are not affected. For such a system, the Hamiltonian is given by
		\begin{equation}
		H(\bm{k})=
		\begin{pmatrix}
		\bm{m}\cdot\bm{\sigma}	& h(\bm{k}) \\
		h^{*}(\bm{k})	& -\bm{m}\cdot\bm{\sigma}
		\end{pmatrix}.
		\end{equation}
		$H_{11}(\bm{k})$, $H_{22}(\bm{k})$ are constants while $H_{12}(\bm{k})$ and $H_{21}(\bm{k})$ are momentum dependent. Then the effective Hamiltonian for the $i=1$ sublattice reads
		\begin{align}\label{eff1}
		H_{\textrm{eff}}^{i=1}(\bm{k})&=\bm{m}\cdot\bm{\sigma}+|h(\bm{k})|^{2}[\varepsilon\sigma_{0}+\bm{m}\cdot\bm{\sigma}]^{-1}\nonumber\\
		&=\bm{m}\cdot\bm{\sigma}+\frac{|h(\bm{k})|^{2}}{m^{2}-\varepsilon^{2}}[\bm{m}\cdot\bm{\sigma}-\varepsilon\sigma_{0}].
		\end{align}
		From Eq (\ref{eff1}), we obtain $\bm{R}^{i=1}(\bm{k},\varepsilon)=\left[1+|h(\bm{k})|^{2}/(m^{2}-\varepsilon^{2})\right]\bm{m}\cdot\bm{\sigma}$. One can easily check that despite the strength of the effective Zeeman field might change depending on the momentum, its direction always remains the same, parallel to the magnetic ordering.
		
		\textit{Non-collinear antiferromagnets on two-dimensional square lattices}.---
		\begin{equation}
		H(\bm{k})=
		\begin{pmatrix}
		\bm{m}_{1}\cdot\bm{\sigma}	& h(\bm{k}) \\
		h^{*}(\bm{k})	& \bm{m}_{2}\cdot\bm{\sigma}
		\end{pmatrix}.
		\end{equation}
		\begin{align}\label{eff2}
		H_{\textrm{eff}}^{i=1}(\bm{k})&=\bm{m}_{1}\cdot\bm{\sigma}+|h(\bm{k})|^{2}[\varepsilon\sigma_{0}-\bm{m}_{2}\cdot\bm{\sigma}]^{-1}\nonumber\\
		&=\bm{m}_{1}\cdot\bm{\sigma}-\frac{|h(\bm{k})|^{2}}{m_{2}^{2}-\varepsilon^{2}}[\varepsilon\sigma_{0}+\bm{m}_{2}\cdot\bm{\sigma}].
		\end{align}
		From Eq (\ref{eff2}), we obtain $\bm{R}^{i=1}(\bm{k},\varepsilon)=\left[\bm{m}_{1}+\bm{m}_{2}|h(\bm{k})|^{2}/(m_{2}^{2}-\varepsilon^{2})\right]\cdot\bm{\sigma}$. It varies its direction by changing $\bm{k}$, but the direction has to be fixed on Fermi surfaces since $|h(\bm{k})|^{2}$ is constant on the Fermi surfaces.
		
		\textit{Non-collinear antiferromagnets on two-dimensional kagome lattices}.---
		When $n$ gets larger than 2, we need to solve more complicated equations to get the effective Hamiltonians.
		three sublattices in one unit cell.
		\begin{align}
		&H(\bm{k})=
		\begin{pmatrix}
		H_{11}(\bm{k})	& H_{12}(\bm{k})	& H_{13}(\bm{k}) \\
		H_{21}(\bm{k})	& H_{22}(\bm{k})	& H_{23}(\bm{k}) \\
		H_{31}(\bm{k})	& H_{32}(\bm{k})	& H_{33}(\bm{k})
		\end{pmatrix} \\
		&G(\bm{k},\varepsilon)=[\varepsilon\tau_{0}\otimes\sigma_{0}-H(\bm{k})]^{-1}=
		\begin{pmatrix}
		G_{11}(\bm{k})	& G_{12}(\bm{k})	& G_{13}(\bm{k}) \\
		G_{21}(\bm{k})	& G_{22}(\bm{k})	& G_{23}(\bm{k}) \\
		G_{31}(\bm{k})	& G_{32}(\bm{k})	& G_{33}(\bm{k})
		\end{pmatrix}
		\end{align}
		
		\begin{align}
		I_{6\times6}&=[\varepsilon\tau_{0}\otimes\sigma_{0}-H(\bm{k})]G(\bm{k},\varepsilon)\nonumber\\
		&=
		\begin{pmatrix}
		[\varepsilon-H_{11}]G_{11}-H_{12}G_{21}-H_{13}G_{31}	& [\varepsilon-H_{11}]G_{12}-H_{12}G_{22}-H_{13}G_{32}	& [\varepsilon-H_{11}]G_{13}-H_{12}G_{23}-H_{13}G_{33} \\
		[\varepsilon-H_{22}]G_{21}-H_{21}G_{11}-H_{23}G_{31}	& [\varepsilon-H_{22}]G_{22}-H_{21}G_{12}-H_{23}G_{32}	& [\varepsilon-H_{22}]G_{23}-H_{21}G_{13}-H_{23}G_{33} \\
		[\varepsilon-H_{33}]G_{31}-H_{31}G_{11}-H_{32}G_{21}	& [\varepsilon-H_{33}]G_{32}-H_{31}G_{12}-H_{32}G_{22}	& [\varepsilon-H_{33}]G_{33}-H_{31}G_{13}-H_{32}G_{23}
		\end{pmatrix}
		\end{align}
		
		\begin{align}
		[\varepsilon-H_{11}]G_{11}-H_{12}G_{21}-H_{13}G_{31}&=I_{2\times2}\nonumber\\
		[\varepsilon-H_{22}]G_{21}-H_{21}G_{11}-H_{23}G_{31}&=0\nonumber\\
		[\varepsilon-H_{33}]G_{31}-H_{31}G_{11}-H_{32}G_{21}&=0
		\end{align}
		
		\begin{align}
		[\varepsilon-H_{11}]G_{11}&=I_{2\times2}+H_{12}G_{21}+H_{13}G_{31}\nonumber\\
		G_{21}&=[\varepsilon-H_{22}]^{-1}[H_{21}G_{11}+H_{23}G_{31}]\nonumber\\
		G_{31}&=[\varepsilon-H_{33}]^{-1}[H_{31}G_{11}+H_{32}G_{21}]
		\end{align}
		
		\begin{align}
		[\varepsilon-H_{11}]G_{11}&=I_{2\times2}+H_{12}G_{21}+H_{13}G_{31}\nonumber\\
		G_{21}&=[\varepsilon-H_{22}]^{-1}[H_{21}G_{11}+H_{23}[\varepsilon-H_{33}]^{-1}[H_{31}G_{11}+H_{32}G_{21}]]\nonumber\\
		G_{31}&=[\varepsilon-H_{33}]^{-1}[H_{31}G_{11}+H_{32}[\varepsilon-H_{22}]^{-1}[H_{21}G_{11}+H_{23}G_{31}]]
		\end{align}
		
		\begin{align}
		[\varepsilon-H_{11}]G_{11}&=I_{2\times2}+H_{12}G_{21}+H_{13}G_{31}\nonumber\\
		[I_{2\times2}-[\varepsilon-H_{22}]^{-1}H_{23}[\varepsilon-H_{33}]^{-1}H_{32}]G_{21}&=[\varepsilon-H_{22}]^{-1}[H_{21}+H_{23}[\varepsilon-H_{33}]^{-1}H_{31}]G_{11}\nonumber\\
		[I_{2\times2}-[\varepsilon-H_{33}]^{-1}H_{32}[\varepsilon-H_{22}]^{-1}H_{23}]G_{31}&=[\varepsilon-H_{33}]^{-1}[H_{31}+H_{32}[\varepsilon-H_{22}]^{-1}H_{21}]G_{11}
		\end{align}
		
		\begin{equation}
		[\varepsilon-H_{11}]G_{11}=I_{2\times2}+H_{12}\frac{[\varepsilon-H_{22}]^{-1}[H_{21}+H_{23}[\varepsilon-H_{33}]^{-1}H_{31}]}{I_{2\times2}-[\varepsilon-H_{22}]^{-1}H_{23}[\varepsilon-H_{33}]^{-1}H_{32}}G_{11}+H_{13}\frac{[\varepsilon-H_{33}]^{-1}[H_{31}+H_{32}[\varepsilon-H_{22}]^{-1}H_{21}]}{I_{2\times2}-[\varepsilon-H_{33}]^{-1}H_{32}[\varepsilon-H_{22}]^{-1}H_{23}}G_{11}
		\end{equation}
		
		\begin{equation}
		\left[\varepsilon-H_{11}-H_{12}\frac{[\varepsilon-H_{22}]^{-1}[H_{21}+H_{23}[\varepsilon-H_{33}]^{-1}H_{31}]}{I_{2\times2}-[\varepsilon-H_{22}]^{-1}H_{23}[\varepsilon-H_{33}]^{-1}H_{32}}-H_{13}\frac{[\varepsilon-H_{33}]^{-1}[H_{31}+H_{32}[\varepsilon-H_{22}]^{-1}H_{21}]}{I_{2\times2}-[\varepsilon-H_{33}]^{-1}H_{32}[\varepsilon-H_{22}]^{-1}H_{23}}\right]G_{11}=I_{2\times2}
		\end{equation}
		
		Since $G_{11}=[\varepsilon-H_{\textrm{eff}}^{i=1}]^{-1}$, we finally obtain
		\begin{equation}
		H_{\textrm{eff}}^{i=1}=H_{11}+H_{12}\frac{[\varepsilon-H_{22}]^{-1}[H_{21}+H_{23}[\varepsilon-H_{33}]^{-1}H_{31}]}{I_{2\times2}-[\varepsilon-H_{22}]^{-1}H_{23}[\varepsilon-H_{33}]^{-1}H_{32}}+H_{13}\frac{[\varepsilon-H_{33}]^{-1}[H_{31}+H_{32}[\varepsilon-H_{22}]^{-1}H_{21}]}{I_{2\times2}-[\varepsilon-H_{33}]^{-1}H_{32}[\varepsilon-H_{22}]^{-1}H_{23}}.
		\end{equation}
		Given that $H_{11}=\bm{m}_{1}\cdot\bm{\sigma}$, $H_{22}=\bm{m}_{2}\cdot\bm{\sigma}$, $H_{33}=\bm{m}_{3}\cdot\bm{\sigma}$, $H_{12}=H_{21}^{\dagger}=h_{12}$, $H_{13}=H_{31}^{\dagger}=h_{13}$, and $H_{23}=H_{32}^{\dagger}=h_{23}$, we find that
		\begin{align}
		H_{\textrm{eff}}^{l=1}=\bm{m}_{1}\cdot\bm{\sigma}&-\frac{|h_{12}|^{2}(m_{3}^{2}-\varepsilon^{2})[\varepsilon+\bm{m}_{2}\cdot\bm{\sigma}]+h_{12}h_{23}h_{31}[\varepsilon+\bm{m}_{2}\cdot\bm{\sigma}][\varepsilon+\bm{m}_{3}\cdot\bm{\sigma}]}{I_{2\times2}-|h_{23}|^{2}[\varepsilon+\bm{m}_{2}\cdot\bm{\sigma}][\varepsilon+\bm{m}_{3}\cdot\bm{\sigma}]}\nonumber\\
		&-\frac{|h_{13}|^{2}(m_{2}^{2}-\varepsilon^{2})[\varepsilon+\bm{m}_{3}\cdot\bm{\sigma}]+h_{13}h_{32}h_{21}[\varepsilon+\bm{m}_{3}\cdot\bm{\sigma}][\varepsilon+\bm{m}_{2}\cdot\bm{\sigma}]}{I_{2\times2}-|h_{23}|^{2}[\varepsilon+\bm{m}_{3}\cdot\bm{\sigma}][\varepsilon+\bm{m}_{2}\cdot\bm{\sigma}]}.
		\end{align}
		Note that $I_{2\times2}-|h_{23}|^{2}[\varepsilon+\bm{m}_{2}\cdot\bm{\sigma}][\varepsilon+\bm{m}_{3}\cdot\bm{\sigma}]=I_{2\times2}-|h_{23}|^{2}[\varepsilon^{2}+\varepsilon\bm{m}_{2}\cdot\bm{\sigma}+\varepsilon\bm{m}_{3}\cdot\bm{\sigma}+\bm{m}_{2}\cdot\bm{m}_{3}+i(\bm{m}_{2}\times\bm{m}_{3})\cdot\bm{\sigma}]$, thus $[I_{2\times2}-|h_{23}|^{2}[\varepsilon+\bm{m}_{2}\cdot\bm{\sigma}][\varepsilon+\bm{m}_{3}\cdot\bm{\sigma}]]^{-1}=[1-|h_{23}|^{2}(\varepsilon^{2}+\bm{m}_{2}\cdot\bm{m}_{3})-|h_{23}|^{2}[\varepsilon(\bm{m}_{2}+\bm{m}_{3})+i(\bm{m}_{2}\times\bm{m}_{3})]\cdot\bm{\sigma}]^{-1}$.
		For simplicity, we define $f=(\varepsilon^{2}+\bm{m}_{2}\cdot\bm{m}_{2})$ and $\bm{g}=\varepsilon(\bm{m}_{2}+\bm{m}_{3})+i(\bm{m}_{2}\times\bm{m}_{3})$. Then we can rewrite the matrix inversion equation as $[(1-|h_{23}|^{2}f)-|h_{23}|^{2}\bm{g}\cdot\bm{\sigma}]^{-1}=[(1-|h_{23}|^{2}f)+|h_{23}|^{2}\bm{g}\cdot\bm{\sigma}]/[(1-|h_{23}|^{2}f)^{2}-|h_{23}|^{4}\bm{g}\cdot\bm{g}]$. Likewise, $[I_{2\times2}-|h_{23}|^{2}[\varepsilon+\bm{m}_{3}\cdot\bm{\sigma}][\varepsilon+\bm{m}_{2}\cdot\bm{\sigma}]]^{-1}=[(1-|h_{23}|^{2}f)+|h_{23}|^{2}\bm{g}^{*}\cdot\bm{\sigma}]/[(1-|h_{23}|^{2}f)^{2}-|h_{23}|^{4}\bm{g}^{*}\cdot\bm{g}^{*}]$
		
		\begin{align}\label{eff3}
		H_{\textrm{eff}}^{i=1}&=\bm{m}_{1}\cdot\bm{\sigma}-[(1-|h_{23}|^{2}f)+|h_{23}|^{2}\bm{g}\cdot\bm{\sigma}]\frac{|h_{12}|^{2}(m_{3}^{2}-\varepsilon^{2})[\varepsilon+\bm{m}_{2}\cdot\bm{\sigma}]+h_{12}h_{23}h_{31}[f+\bm{g}\cdot\bm{\sigma}]}{(1-|h_{23}|^{2}f)^{2}-|h_{23}|^{4}\bm{g}\cdot\bm{g}}\nonumber\\
		&\qquad\qquad\,-[(1-|h_{23}|^{2}f)+|h_{23}|^{2}\bm{g}^{*}\cdot\bm{\sigma}]\frac{|h_{13}|^{2}(m_{2}^{2}-\varepsilon^{2})[\varepsilon+\bm{m}_{3}\cdot\bm{\sigma}]+h_{13}h_{32}h_{21}[f+\bm{g}^{*}\cdot\bm{\sigma}]}{(1-|h_{23}|^{2}f)^{2}-|h_{23}|^{4}\bm{g}^{*}\cdot\bm{g}^{*}}\nonumber\\
		&=R^{i=1}_{0}(\bm{k},\varepsilon)+\bm{R}^{i=1}(\bm{k},\varepsilon)\cdot\bm{\sigma}.
		\end{align}
		To acquire a simplified formula for $\bm{R}^{i=1}(\bm{k},\varepsilon)$, we use $(\bm{g}\cdot\bm{\sigma})(\bm{m}_{2}\cdot\bm{\sigma})=\bm{g}\cdot\bm{m}_{2}+i(\bm{g}\times\bm{m}_{2})\cdot\bm{\sigma}=\varepsilon(m_{2}^{2}+\bm{m}_{2}\cdot\bm{m}_{3})+i[\varepsilon\bm{m}_{3}\times\bm{m}_{2}+i(m_{2}^{2}\bm{m}_{3}-(\bm{m}_{2}\cdot\bm{m}_{3})\bm{m}_{2})]\cdot\bm{\sigma}$, $(\bm{g}^{*}\cdot\bm{\sigma})(\bm{m}_{3}\cdot\bm{\sigma})=\bm{g}^{*}\cdot\bm{m}_{3}+i(\bm{g}^{*}\times\bm{m}_{3})\cdot\bm{\sigma}=\varepsilon(m_{3}^{2}+\bm{m}_{2}\cdot\bm{m}_{3})+i[\varepsilon\bm{m}_{2}\times\bm{m}_{3}-i((\bm{m}_{2}\cdot\bm{m}_{3})\bm{m}_{3}-m_{3}^{2}\bm{m}_{2})]\cdot\bm{\sigma}$, $(\bm{g}\cdot\bm{\sigma})(\bm{g}\cdot\bm{\sigma})=\bm{g}\cdot\bm{g}+i(\bm{g}\times\bm{g})\cdot\bm{\sigma}=\bm{g}\cdot\bm{g}$ and $(\bm{g}^{*}\cdot\bm{\sigma})(\bm{g}^{*}\cdot\bm{\sigma})=\bm{g}^{*}\cdot\bm{g}^{*}+i(\bm{g}^{*}\times\bm{g}^{*})\cdot\bm{\sigma}=\bm{g}^{*}\cdot\bm{g}^{*}$ ($\bm{g}\cdot\bm{g}=\bm{g}^{*}\cdot\bm{g}^{*}=\varepsilon^{2}(\bm{m}_{2}+\bm{m}_{3})^{2}-(\bm{m}_{2}\times\bm{m}_{3})\cdot(\bm{m}_{2}\times\bm{m}_{3})$). Then the numerators of Eq (\ref{eff3}) read
		\begin{align}
		-[\varepsilon\bm{g}\cdot\bm{\sigma}+(\bm{g}\cdot\bm{\sigma})(\bm{m}_{2}\cdot\bm{\sigma})]&=-[\cdots+\varepsilon^{2}(\bm{m}_{2}+\bm{m}_{3})+i\varepsilon(\bm{m}_{2}\times\bm{m}_{3})+i[\varepsilon\bm{m}_{3}\times\bm{m}_{2}+i(m_{2}^{2}\bm{m}_{3}-(\bm{m}_{2}\cdot\bm{m}_{3})\bm{m}_{2})]]\cdot\bm{\sigma}\nonumber\\
		&=-[\cdots+\varepsilon^{2}(\bm{m}_{2}+\bm{m}_{3})-(m_{2}^{2}\bm{m}_{3}-(\bm{m}_{2}\cdot\bm{m}_{3})\bm{m}_{2})]\cdot\bm{\sigma}\nonumber\\
		-[\varepsilon\bm{g}^{*}\cdot\bm{\sigma}+(\bm{g}^{*}\cdot\bm{\sigma})(\bm{m}_{2}\cdot\bm{\sigma})]&=-[\cdots+\varepsilon^{2}(\bm{m}_{2}+\bm{m}_{3})-i\varepsilon(\bm{m}_{2}\times\bm{m}_{3})+i[\varepsilon\bm{m}_{2}\times\bm{m}_{3}-i((\bm{m}_{2}\cdot\bm{m}_{3})\bm{m}_{3}-m_{3}^{2}\bm{m}_{2})]]\cdot\bm{\sigma}\nonumber\\
		&=-[\cdots+\varepsilon^{2}(\bm{m}_{2}+\bm{m}_{3})+((\bm{m}_{2}\cdot\bm{m}_{3})\bm{m}_{3}-m_{3}^{2}\bm{m}_{2})]\cdot\bm{\sigma}\nonumber\\
		&\vdots
		\end{align}
		where the terms which belong to $R^{i=1}_{0}(\bm{k},\varepsilon)$ part are omitted. The denominators read $(1-|h_{23}|^{2}f)^{2}-|h_{23}|^{4}\bm{g}\cdot\bm{g}=(1-|h_{23}|^{2}f)^{2}-|h_{23}|^{4}\bm{g}^{*}\cdot\bm{g}^{*}=(1-|h_{23}|^{2}f)^{2}-|h_{23}|^{4}(\bm{m}_{2}+\bm{m}_{3})^{2}$. Finally, we obtain
		\begin{align}\label{eff4}
		\bm{R}^{i=1}(\bm{k},\varepsilon)\cdot\bm{\sigma}&=\bm{m}_{1}\cdot\bm{\sigma}-|h_{12}|^{2}(m_{3}^{2}-\varepsilon^{2})\frac{[(1-|h_{23}|^{2}f)\bm{m}_{2}+|h_{23}|^{2}[\varepsilon^{2}(\bm{m}_{2}+\bm{m}_{3})-m_{2}^{2}\bm{m}_{3}+(\bm{m}_{2}\cdot\bm{m}_{3})\bm{m}_{2}]]\cdot\bm{\sigma}}{(1-|h_{23}|^{2}f)^{2}-|h_{23}|^{4}[\varepsilon^{2}(\bm{m}_{2}+\bm{m}_{3})^{2}-(\bm{m}_{2}\times\bm{m}_{3})\cdot(\bm{m}_{2}\times\bm{m}_{3})]}\nonumber\\
		&\qquad\qquad\,-|h_{13}|^{2}(m_{2}^{2}-\varepsilon^{2})\frac{[(1-|h_{23}|^{2}f)\bm{m}_{3}+|h_{23}|^{2}[\varepsilon^{2}(\bm{m}_{2}+\bm{m}_{3})-m_{3}^{2}\bm{m}_{2}+(\bm{m}_{2}\cdot\bm{m}_{3})\bm{m}_{3}]]\cdot\bm{\sigma}}{(1-|h_{23}|^{2}f)^{2}-|h_{23}|^{4}[\varepsilon^{2}(\bm{m}_{2}+\bm{m}_{3})^{2}-(\bm{m}_{2}\times\bm{m}_{3})\cdot(\bm{m}_{2}\times\bm{m}_{3})]}\nonumber\\
		&\qquad\qquad\,-\frac{[(1-|h_{23}|^{2}f)h_{12}h_{23}h_{31}+|h_{23}|^{2}h_{12}h_{23}h_{31}f]\bm{g}\cdot\bm{\sigma}+\textit{c.c.}}{(1-|h_{23}|^{2}f)^{2}-|h_{23}|^{4}[\varepsilon^{2}(\bm{m}_{2}+\bm{m}_{3})^{2}-(\bm{m}_{2}\times\bm{m}_{3})\cdot(\bm{m}_{2}\times\bm{m}_{3})]}.
		\end{align}
		For the kagome lattice without spin-orbit interaction, we can choose a gauge in which $h_{12}(\bm{k})$, $h_{23}(\bm{k})$, and $h_{31}(\bm{k})$ are real functions due to the symmetries ($C_{2z}$ and $P$) of the lattice. In such a present case, the fourth term of Eq (\ref{eff4}) can be shorten into $2h_{12}h_{23}h_{31}\textrm{Re}(\bm{g})\cdot\bm{\sigma}/[(1-|h_{23}|^{2}f)^{2}-|h_{23}|^{4}(\bm{m}_{2}+\bm{m}_{3})^{2}]$. Collecting together the terms that involving $\bm{m}_{2}$ and $\bm{m}_{3}$, we find
		\begin{align}\label{eff5}
		\bm{R}^{i=1}(\bm{k},\varepsilon)&=\bm{m}_{1}-|h_{12}|^{2}(m_{3}^{2}-\varepsilon^{2})\frac{(1-|h_{23}|^{2}f)\bm{m}_{2}+|h_{23}|^{2}[\varepsilon^{2}(\bm{m}_{2}+\bm{m}_{3})-m_{2}^{2}\bm{m}_{3}+(\bm{m}_{2}\cdot\bm{m}_{3})\bm{m}_{2}]}{(1-|h_{23}|^{2}f)^{2}-|h_{23}|^{4}[\varepsilon^{2}(\bm{m}_{2}+\bm{m}_{3})^{2}-(\bm{m}_{2}\times\bm{m}_{3})\cdot(\bm{m}_{2}\times\bm{m}_{3})]}\nonumber\\
		&\quad\quad\ \ \,-|h_{13}|^{2}(m_{2}^{2}-\varepsilon^{2})\frac{(1-|h_{23}|^{2}f)\bm{m}_{3}+|h_{23}|^{2}[\varepsilon^{2}(\bm{m}_{2}+\bm{m}_{3})-m_{3}^{2}\bm{m}_{2}+(\bm{m}_{2}\cdot\bm{m}_{3})\bm{m}_{3}]}{(1-|h_{23}|^{2}f)^{2}-|h_{23}|^{4}[\varepsilon^{2}(\bm{m}_{2}+\bm{m}_{3})^{2}-(\bm{m}_{2}\times\bm{m}_{3})\cdot(\bm{m}_{2}\times\bm{m}_{3})]}\nonumber\\
		&\quad\quad\ \ \,-\frac{2h_{12}h_{23}h_{31}\varepsilon(\bm{m}_{2}+\bm{m}_{3})\cdot\bm{\sigma}}{(1-|h_{23}|^{2}f)^{2}-|h_{23}|^{4}[\varepsilon^{2}(\bm{m}_{2}+\bm{m}_{3})^{2}-(\bm{m}_{2}\times\bm{m}_{3})\cdot(\bm{m}_{2}\times\bm{m}_{3})]}\nonumber\\
		&=\bm{m}_{1}-\bm{m}_{2}\frac{|h_{12}|^{2}(m_{3}^{2}-\varepsilon^{2})[1+|h_{23}|^{2}(\varepsilon^{2}+\bm{m}_{2}\cdot\bm{m}_{3}-f)]-|h_{13}|^{2}|h_{23}|^{2}(m_{2}^{2}-\varepsilon^{2})(m_{3}^{2}-\varepsilon^{2})+2h_{12}h_{23}h_{31}\varepsilon}{(1-|h_{23}|^{2}f)^{2}-|h_{23}|^{4}[\varepsilon^{2}(\bm{m}_{2}+\bm{m}_{3})^{2}-(\bm{m}_{2}\times\bm{m}_{3})\cdot(\bm{m}_{2}\times\bm{m}_{3})]}\nonumber\\
		&\quad\quad\ \ \,-\bm{m}_{3}\frac{|h_{13}|^{2}(m_{2}^{2}-\varepsilon^{2})[1+|h_{23}|^{2}(\varepsilon^{2}+\bm{m}_{2}\cdot\bm{m}_{3}-f)]-|h_{12}|^{2}|h_{23}|^{2}(m_{2}^{2}-\varepsilon^{2})(m_{3}^{2}-\varepsilon^{2})+2h_{12}h_{23}h_{31}\varepsilon}{(1-|h_{23}|^{2}f)^{2}-|h_{23}|^{4}[\varepsilon^{2}(\bm{m}_{2}+\bm{m}_{3})^{2}-(\bm{m}_{2}\times\bm{m}_{3})\cdot(\bm{m}_{2}\times\bm{m}_{3})]}\nonumber\\
		&=\bm{m}_{1}+\mathfrak{f}(\bm{k})\bm{m}_{2}+\mathfrak{g}(\bm{k})\bm{m}_{3}.
		\end{align}
		In the absence of $C_{2z}$ or $P$ symmetries, $h_{12}(\bm{k})$, $h_{23}(\bm{k})$, and $h_{31}(\bm{k})$ are complex, and $\bm{R}^{i=1}(\bm{k},\varepsilon)$ acquires an additional term $\mathfrak{h}(\bm{k})$ which is proportional to $\textrm{Im}(\bm{g})=\bm{m}_{2}\times\bm{m}_{3}$. This means that even when $\bm{m}_{1}$, $\bm{m}_{2}$, and $\bm{m}_{3}$ are coplanar, the spin texture in the momentum space can have a non-zero out-of-plane component.
		\\
		\\
		\begin{table}[h!]
			\caption{The transformation properties of $A_{1u}$, $A_{2u}$, and $E_u$ representations under the generators of the $D_{3d}$ point group}
			\centering
			\begin{tabular}{|c|c|c|c|c|}
				\hline
				& $E$ & $C_{3z}$ & $C_{2y}$ & $P$ \\ \hline
				$A_{1u}$ & 1 & 1 & 1 & -1 \\ \hline
				$A_{2u}$ & 1 & 1 & -1 & -1 \\ \hline
				$E_{u}$ & 2 & -1 & 0 & -2 \\ \hline
			\end{tabular}
			\label{table1}
		\end{table}
		\noindent
		\textbf{$E_u$ irreducible representation}\label{section3}\\
		In general, a pairing function that belongs to the $E_u$ irreducible representation (IR) can be written as
		\begin{equation}
		\Delta_{\bm{k}}^{E_u}=[\phi_1\bm{d}_1(\bm{k})\cdot\bm{\sigma}+\phi_2\bm{d}_2(\bm{k})\cdot\bm{\sigma}](i\sigma_y),
		\end{equation}
		where $\bm{d}_1(\bm{k})$ and $\bm{d}_2(\bm{k})$ are the two basis functions of the $E_u$ IR whose lowest order terms in $k_x$ and $k_y$ are given by $\bm{d}_1(\bm{k})=k_x\hat{x}$ and $\bm{d}_2(\bm{k})=k_y\hat{y}$.
		A general expression for $C_{3z}$ symmetric Ginzburg-Landau free energy density described by order parameters $\phi_1$ and $\phi_2$ is
		\begin{equation}
		F=\alpha(T-T_c)(|\phi_1|^2+|\phi_2|^2)+\beta_1(|\phi_1|^2+|\phi_2|^2)^2+\beta_2(\phi_1\phi_2^*-\phi_1^*\phi_2)^2.
		\end{equation}
		When the coefficients satisfy $\alpha<0$ at $T<T_c$ and $\beta_1>0$, $F$ has global minima at finite $(\phi_1,\phi_2)$s at temperatures lower than $T_c$.
		If $\beta_2>0$, $(\phi_1,\phi_2)=\phi(1,\pm i)$, and $\Delta_{\bm{k}}^{E_u}$ is symmetric under $C_{3z}$.
		On the other hand, if $\beta_2>0$, $(\phi_1,\phi_2)=\phi(1,\pm 1)$, and $\Delta_{\bm{k}}^{E_u}$ breaks $C_{3z}$ (nematic phase).
		In this letter, to keep $C_{3z}$ symmetry which plays an important role in the gap structure and band topology, we only consider the 
		$\beta_2>0$ case. We note that $\phi(1,i)$ and $\phi(1,-i)$ pairing functions transform to each other under in-plane two-fold rotations.
		Thus, each individual ground state breaks the two-fold rotation symmetries, and does not have a symmetry constraint for its $d$-vector that forces superconducting gap nodes.
		\\
		\\
		\noindent
		\textbf{Topological superconductivity}\label{section4}\\
		\indent The Chern number and the $\mathbb{Z}_4$ invariant that characterize the first-order TSCs and inversion symmetry protected second-order TSCs can be obtained by the symmetry indicator approach~\cite{skurativska2020atomic,ono2020refined,ono2021z,elcoro2021magnetic}. First, in the presence of $C_{3z}$ and $P$ symmetries, the Chern number is given by $\mathcal{C}=-2z_{C_{3z}}^{\text{BdG}}+3z_I^{\text{BdG}}\mod6$, where $z_{C_{3z}}^{\text{BdG}}=\sum_{\bm{k}=\Gamma,K,K'}(q_{\bm{k}}^{-\omega,{\text{BdG}}}-q_{\bm{k}}^{-1,{\text{BdG}}})$ and $z_{P}^{\text{BdG}}=\sum_{\bm{k}=\Gamma,M_{i=1,2,3}}p_{\bm{k}}^{-,{\text{BdG}}}$. $q_{\bm{k}}^{\lambda,{\text{BdG}}}$ ($p_{\bm{k}}^{\nu,{\text{BdG}}}$) is the number of occupied states of $\mathcal{H}_{\text{BdG}}(\bm{k})$ with $C_{3z}$ ($P$) eigenvalue $\lambda$ ($\nu$) at a $C_{3z}$ ($P$) invariant point $\bm{k}=\Gamma,K,K'$ ($\bm{k}=\Gamma,M_{i=1,2,3}$). Second, the $\mathbb{Z}_4$ invariant is given by $\mathcal{Z}=\sum_{\bm{k}=\Gamma,M_{i=1,2,3}}\sum_{\nu}(\nu/2)(p_{\bm{k}}^{\nu,{\text{BdG}}}-p_{\bm{k}}^{\nu,{\text{BdG}}}|_{\mu=-\infty})\mod4$.
		
		The above equations for topological invariants are expressed in terms of $q_{\bm{k}}^{\lambda,{\text{BdG}}}$ and $p_{\bm{k}}^{\nu,{\text{BdG}}}$. In the weak-coupling limit, they can be directly obtained from the normal state.
		$C_{3z}$, $P$, and particle-hole symmetry yield the following relations:
		\begin{align}\label{inv}
		z_{C_{3z}}^{\text{BdG}}&=
		\begin{cases}
		2q_{K}^{-\omega,{\text{N}}}-2q_{K}^{-\omega^2,{\text{N}}}+q_{\Gamma}^{-\omega,{\text{N}}}-q_{\Gamma}^{-\omega^2,{\text{N}}}&A_{1u}\\
		4q_{K}^{-\omega,{\text{N}}}-4q_{K}^{-1,{\text{N}}}+2q_{\Gamma}^{-\omega,{\text{N}}}-2q_{\Gamma}^{-1,{\text{N}}}&E_{u,\alpha=\omega}\\
		2q_{K}^{-\omega^2,{\text{N}}}-2q_{K}^{-1,{\text{N}}}+q_{\Gamma}^{-\omega^2,{\text{N}}}-q_{\Gamma}^{-1,{\text{N}}}&E_{u,\alpha=\omega^2}\\
		\end{cases}\nonumber\\
		&~\mod 3,\nonumber\\
		z_{P}^{\text{BdG}}&=p_{\Gamma}^{-,{\text{N}}}+p_{\Gamma}^{+,{\text{N}}}+3(p_{M_1}^{-,{\text{N}}}+p_{M_1}^{+,{\text{N}}})\mod 2,\nonumber\\
		\mathcal{Z}&=p_{\Gamma}^{+,{\text{N}}}-p_{\Gamma}^{-,{\text{N}}}+3(p_{M_1}^{+,{\text{N}}}-p_{M_1}^{-,{\text{N}}})\mod 4,
		\end{align}
		where $q_{\bm{k}}^{\lambda,{\text{N}}}$ ($p_{\bm{k}}^{\nu,{\text{N}}}$) is the number of occupied states of the normal state Hamiltonian $\mathcal{H}_{\text{KAFM}}(\bm{k})$ with $C_{3z}$ ($P$) eigenvalue $\lambda$ ($\nu$) at a $C_{3z}$ ($P$) invariant point.
		$\alpha$ is a $C_{3z}$ eigenvalue of $\Delta_{\bm{k}}^{E_u}$ ($C_{3z}(\bm{k})\Delta_{\bm{k}}^{E_u}(\bm{k})C_{3z}^T(\bm{k})=\alpha\Delta_{C_{3z}\bm{k}}^{E_u}$ and $\omega\equiv e^{2i\pi/3}$).
		
		\begin{figure}[t!]
			\centering
			\includegraphics[width=0.6\linewidth]{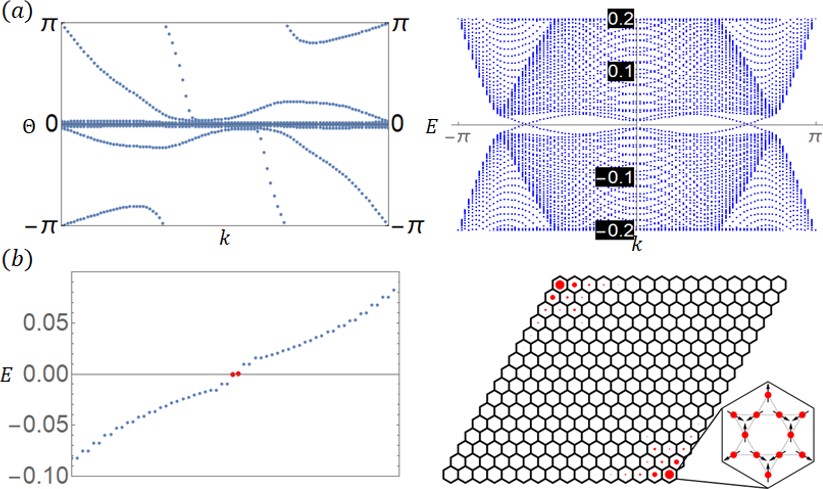}
			\caption{
				(a) The Wilson loop eigenvalue spectrum of the occupied states along the $k_x$-direction (left), and the energy spectrum of a tight-binding model in a strip geometry periodic along the $x$-direction while finite along the $y$-direction (right) of $\mathcal{H}_{\text{BdG}}(\bm{k})$ with $\Delta_{\bm{k}}^{E_{u}}$, $m=0.2$ and $\mu=-3.5$. In accord with Fig. 3 (b), winding in the Wilson loop spectrum shows that $\mathcal{C}=2$ in this condition. Corresponding to the non-trivial bulk topology, two chiral edge modes appear at each edge.
				(b) The Wilson loop spectrum (left) and the zero-energy Majorana modes at the corners of a finite-size system cut into an inversion symmetric geometry (right) of $\mathcal{H}_{\text{BdG}}(\bm{k})$ with $\Delta_{\bm{k}}^{A_{1u}}$, $m=0.2$ and $\mu=-3.5$. Here, $\mathcal{C}=0$, so the edge modes are gapped. However the Majorana corner modes appear, since $\mathcal{Z}=1$. The radii of the red dots in the right panel represent the weight of the unit cells at their positions in the corner mode wavefunction.
			}\label{fig4}
		\end{figure}
		
		When $\mathcal{C}\neq0$, the system becomes a first-order TSC. As shown in Fig.~\ref{fig4} (a), $\mathcal{C}$ chiral edge modes appear on the boundaries of the first-order TSC~\cite{qi2006general,fu2010odd,sato2010topological,fukui2012bulk,chiu2016classification}. When $\mathcal{C}=0$ and $\mathcal{Z}=1$, the system becomes a second-order TSC protected by inversion symmetry in which the edge spectrum is gapped as in Fig.~\ref{fig4} (b)~\cite{khalaf2018higher,ahn2020higher,skurativska2020atomic,ghorashi2023altermagnetic}. We note that if the pairing interaction preserves $C_{2z}T$ symmetry, a combination of a two-fold rotation around the $z$-axis $C_{2z}$ and time-reversal operation $T$ that satisfies $(C_{2z}T)^2=1$, one can choose a gauge in which $\mathcal{H}_{\text{BdG}}(\bm{k})$ is real. Then $\mathcal{H}_{\text{BdG}}(\bm{k})$ can be classified as a Stiefel-Whitney superconductor, whose topology is characterized by the second Stiefel-Whitney number ($\mathbb{Z}_2$). The second Stiefel-Whitney number can be detected by counting the parity of Wilson band crossings at the $\Theta=\pi$ line (See Fig.~\ref{fig4} (b)). Cut into a finite-size inversion symmetric geometry, the second-order TSC hosts zero-energy Majorana modes localized at the corners (Fig.~\ref{fig4} (b)).

	\end{bibunit}
\end{document}